\documentclass[12pt,thmsa]{article}
\usepackage{sw20lart}

\input tcilatex
\QQQ{Language}{
American English
}

\begin{document}

\author{I. Radinschi \and ''Gh. Asachi'' Technical University, Department of
Physics, \and B-dul Dimitrie Mangeron, No. 67, Iasi, Romania, 6600 \and %
iradinsc@phys.tuiasi.ro}
\title{The Energy Distribution in a Static Spherically Symmetric Nonsingular Black
Hole Space-Time}
\date{May 17, 2000 }
\maketitle

\begin{abstract}
We calculate the energy distribution in a static spherically symmetric
nonsingular black hole space-time by using the Tolman's energy-momentum
complex. All the calculations are performed in quasi-Cartesian coordinates.
The energy distribution is positive everywhere and be equal to zero at
origin. We get the same result as obtained by Y-Ching Yang by using the
Einstein's and Weinberg's prescriptions.


\end{abstract}

\section{INTRODUCTION}

The General Theory of Relativity is an excellent theory of space, time and
gravitation and has been supported by experimental evidences. However, the
subject of energy-momentum localization has been a problematic issue since
the outset of this theory.

A large number of definitions of the gravitational energy have been given
since now. Some of them are coordinate independent and other are
coordinate-dependent. It is possible to evaluate the energy and momentum
distribution by using various energy-momentum complexes. There lies a
dispute with the importance of nontensorial energy-momentum complexes whose
physical interpretation have been questioned by a number of physicists,
including Weyl, Pauli and Eddington. Also, there exist an opinion that the
energy-momentum complexes are not useful to get meaningful energy
distribution in a given geometry.

Several examples of particular space-times (the Kerr-Newman, the
Einstein-Rosen and the Bonnor-Vaidya) have been investigated and different
energy-momentum complexes are known to give the same energy distribution for
a given space-time [1]-[6]. Aguirregabiria, Chamorro and Virbhadra [7]
showed that several energy-momentum complexes coincide for any Kerr-Schild
class metric. Xulu obtained interesting results about the energy
distribution of a charged dilaton black hole [8] and about the energy
associated with a Schwarzschild black hole in a magnetic universe [9]. Also,
recently, Xulu [10] obtained the total energy of a model of universe based
on the Bianchi I type metric. Recently, Virbhadra [11] shows that different
energy-momentum complexes give the same and reasonable results for many
space-times.

I. C. Yang [12] employing the Einstein's and Weinberg's energy-momentum
complexes obtained the energy distribution in the Dymnikova space-time that
is positive everywhere and be equal to zero at origin.

In this paper we compute the energy in a static spherically symmetric
nonsingular black hole space-time in the Tolman's prescription [13] . We
obtain the same result as obtained by I. Ching-Yang [12] and also make a
discussion of the results. We use the geometrized units $(G=1,c=1)$ and
follow the convention that the Latin indices run from $0$ to $3$.

\section{THE\ ENERGY\ DISTRIBUTION IN\ THE\ TOLMAN'S\ PRESCRIPTION}

Dymnikova [14] obtained a static spherically symmetric nonsingular black
hole solution which is expressed by

\begin{equation}
ds^2=(1-\frac{R_g(r)}r)dt^2-\frac{dr^2}{(1-\frac{R_g(r)}r)}-r^2d\theta
^2-r^2\sin \theta ^2d\varphi ^2,
\end{equation}

where

\begin{equation}
R_g(r)=r_g(1-e^{(\frac{r^3}{r_1^3})}),
\end{equation}

and 
\[
\]
\begin{equation}
r_1^3=r_0^2r_g.
\end{equation}
We have also

\begin{equation}
r_0^2=\frac 3{8\pi \varepsilon _0},
\end{equation}
and

\begin{equation}
r_g=2M.
\end{equation}
This black hole solution is regular at $r=0$ and everywhere else. The
assumed form of the energy-momentum tensor is

\begin{equation}
T_0^0=\varepsilon _0e^{-\frac{r3}{r_0^2r_g}}.
\end{equation}

The structure of this solution is like a Schwarzschild solution whose
singularity is replaced by the de Sitter core.

The Tolman's energy-momentum complex [13] is given by

\begin{equation}
\Upsilon _i^k=\frac 1{8\pi }U_i^{kl},_l,
\end{equation}

where $\Upsilon _0^0$ and $\Upsilon _\alpha ^0$ are the energy and momentum
components.

We have

\begin{equation}
U_i^{kl}=\sqrt{-g}(-g^{pk}V_{ip}^l+\frac 12g_i^kg^{pm}V_{pm}^l),
\end{equation}

with

\begin{equation}
V_{jk}^i=-\Gamma _{jk}^i+\frac 12g_j^i\Gamma _{mk}^m+\frac 12g_k^i\Gamma
_{mj}^m.
\end{equation}

The energy-momentum complex $\Upsilon _i^k$ also satisfies the local
conservation laws

\begin{equation}
\frac{\partial \Upsilon _i^k}{\partial x^k}=0.
\end{equation}

The Tolman's energy-momentum complex gives the correct result if the
calculations are carried out in quasi-Cartesian coordinates.

We transform the line element (1) to quasi-Cartesian coordinates $t,x,y,z$
according to

\begin{equation}
\begin{tabular}{c}
$x=r\sin \theta \cos \varphi ,$ \\ 
$y=r\sin \theta \sin \varphi ,$ \\ 
$z=r\cos \theta ,$%
\end{tabular}
\end{equation}

and

\begin{equation}
r=(x^2+y^2+z^2)^{\frac 12}.
\end{equation}

The line element (1) becomes

\begin{equation}
ds^2=(1-\frac{R_g(r)}r)dt^2-(dx^2+dy^2+dz^2)-\frac{R_g(r)}{r^2(r-R_g(r))}%
(xdx+ydy+zdz)^2.
\end{equation}

The only required components of $U_i^{kl}$ in the calculation of the energy
are the following

\[
U_0^{01}=\frac{xR_g(r)}{r^3}, 
\]

\begin{equation}
U_0^{02}=\frac{yR_g(r)}{r^3},
\end{equation}

\[
U_0^{03}=\frac{zR_g(r)}{r^3}. 
\]

The components of $U_i^{kl}$ are calculated with the program Maple GR Tensor
II Release 1.50.

The energy and momentum in the Tolman's prescription are given by

\begin{equation}
P_i=\iiint \Upsilon _i^0dx^1dx^2dx^3.
\end{equation}

Using the Gauss's theorem we obtain

\begin{equation}
P_i=\frac 1{8\pi }\iint U_i^{0\alpha }n_\alpha dS
\end{equation}

where $n_\alpha =(x/r,y/r,z/r)$ are the components of a normal vector over
an infinitesimal surface element $dS=r^2\sin \theta d\theta d\varphi $.

Using (14) and applying the Gauss's theorem (16) we evaluate the integral
over the surface of a sphere with radius $r$

\begin{equation}
E(r)=\frac 1{8\pi }\iint \frac{R_g(r)}{r^2}r^2\sin \theta d\theta d\varphi .
\end{equation}

We find that the energy within a sphere with radius $r$ is given

\begin{equation}
E(r)=\frac{r_g}2(1-e^{(-\frac{r^3}{r_1^3})}).
\end{equation}

As $r\rightarrow 0$, $E(r)\rightarrow 0$ and as $r\rightarrow \infty $, $%
E(r)\rightarrow M$. Thus the total energy is given by the parameter $M$
which is the same as the ADM mass for this metric. Note that $E(r)>0$ for
all $r$: $0\leq r<\infty $.

\section{DISCUSSION}

The subject of the localization of energy continues to be an interesting
one. Bondi [15] sustained that a nonlocalizable form of energy is not
admissible in relativity. Other authors consider that because the
energy-momentum complexes are not tensorial objects and give results which
are coordinate dependent they are not adequate for describing the
gravitational field.

Due to this reason, this subject was not take in seriously for a long time
and was re-opened by the results obtained by Virbhadra, Rosen, Chamorro and
Aguirregabiria. Some interesting results which have been found recently
[1]-[6], [7]-[12] lend support to the idea that the several energy-momentum
complexes can give the same and acceptable result for a given space-time.
Also, in his recent paper Virbhadra [11] emphasized that though the
energy-momentum complexes are non-tensors under general coordinate
transformations, the local conservation laws with them hold in all
coordinate systems. Chang, Nester and Chen [16] showed that there exist a
direct relationship between energy-momentum complexes and quasilocal
expressions because every energy-momentum complexes is associated with a
legitimate Hamiltonian boundary term.

Our result also lends support to the idea that the energy-momentum complexes
can give the same result for a given space-time.

\end{document}